\documentclass[twocolumn,aip,apl,reprint,showpacs,preprintnumbers,amsmath,amssymb,floatfix,superscriptaddress]{revtex4-1}
\usepackage{graphicx}
\usepackage{amsmath}
\usepackage{amssymb}
\usepackage{color}

\begin{document}

\title{Electrical transport in onion-like carbon - PMMA nanocomposites}

\author{Claudio Grimaldi}\email{claudio.grimaldi@epfl.ch}\affiliation{Laboratory of Physics of Complex Matter, Ecole Polytechnique F\'ed\'erale de Lausanne, Station 3, CH-1015 Lausanne, Switzerland}
\author{Egon Kecsenovity}\affiliation{Department of Physical Chemistry and Materials Science, University of Szeged, Rerrich Square 1, Szeged H-6720, Hungary}
\author{Maryam Majidian}\affiliation{Laboratory of Physics of Complex Matter, Ecole Polytechnique F\'ed\'erale de Lausanne, Station 3, CH-1015 Lausanne, Switzerland}
\author{Vladimir L. Kuznetsov}\affiliation{Boreskov Institute of Catalysis, SB RAS, Lavrentieva 5, Novosibirsk, 630090, Russia}
\author{Arnaud Magrez}\affiliation{Laboratory of Physics of Complex Matter, Ecole Polytechnique F\'ed\'erale de Lausanne, Station 3, CH-1015 Lausanne, Switzerland}
\affiliation{Crystal Growth Facility, Ecole Polytechnique F\'ed\'erale de Lausanne, Station 3, CH-1015 Lausanne, Switzerland}
\author{L\'aszl\'o Forr\'o}\affiliation{Laboratory of Physics of Complex Matter, Ecole Polytechnique F\'ed\'erale de Lausanne, Station 3, CH-1015 Lausanne, Switzerland}

\begin{abstract}
We report electrical conductivity measurements of Polymethyl-methacrylate filled by onion-like carbon particles with primary particle size of $\approx 5$ nm.
We shown that the conductivity $\sigma$ is exceptionally high even at very low loadings, and that its low-temperature dependence follows
a Coulomb gap regime at atmospheric pressure and an activated behavior at a pressure of $2$ GPa. We interpret this finding in terms of 
the enhancement under the applied pressure of the effective dielectric permittivity within the aggregates of onion-like carbons, which improves the 
screening of the Coulomb interaction and reduces the optimal hopping distance of the electrons.

\end{abstract}

\maketitle 

Onion-like carbon (OLC) nanostructures, also called carbon onions, are nanometric quasi-spherical 
particles consisting of multi-shell graphitic layers, obtained by the transformation of ultra-dispersed 
nano-diamonds through annealing above $1200$ K.\cite{Kuznetsov1994} When dispersed in 
polymeric host materials, the high electrical conductivity of OLC nanoparticles, as well as their lightweight and inert nature,
are attractive properties for electromagnetic shielding applications, which can be controlled by tailoring the OLC
cluster size and nano-diamond annealing temperature.\cite{Kuzhir2012} OLCs are interesting also for high
power applications, due to their high charge-discharge rates, and for energy storage applications, as
OLC particles act effectively as electrical double-layer micro-capacitors.\cite{Ganesh2011}

In addition to the properties specific to OLC particles, equally important factors influencing the performance
of OLC-based composites are the quality and type of the OLCs dispersion in the
host material, the operational temperature, and the response to external fields.
Here, we report on our studies on the filler concentration, pressure, and temperature dependence of the electrical
conductivity of OLC particles dispersed in Polymethyl-methacrylate (PMMA). We show that OLC-PMMA
composites display an exceptionally high value of the conductivity, with no sign of a percolative transition
even for filler concentrations as low as $0.38$ \% in volume fraction of the OLC particles. Even more strikingly, we find that
the low-temperature conductivity crosses over from a Coulomb gap regime at atmospheric pressure to
an activated behavior at a hydrostatic applied pressure of $P=2$ GPa, regardless of the OLC concentrations considered. 
These results point to a non-trivial interplay between the morphology of the OLC aggregates in the polymer and the electron
hopping processes between OLC nanoparticles. 

\begin{figure}[t]
\begin{center}
\includegraphics[scale=1,clip=true]{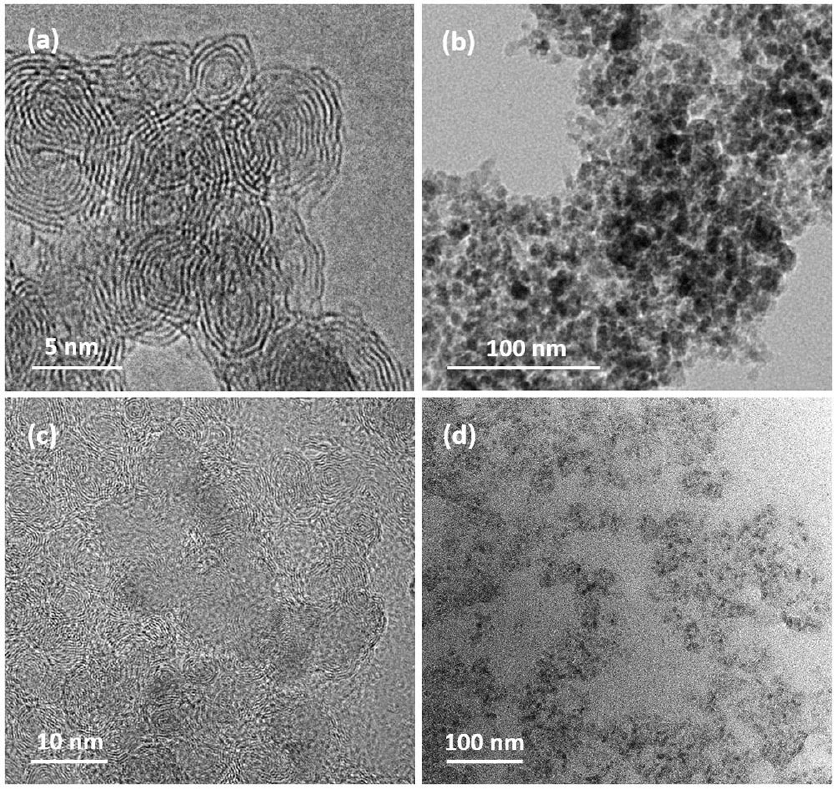}
\caption{TEM images of OLC powders, (a) and (b), and of OLC-PMMA with $1$ wt\% filler loading, (c) and (d). The images in (c) and (d) show that
the dispersion of OLC particles in the PMMA matrix is characterized by aggregates of closely separated OLC fillers, giving rise to a locally non-homogeneous
microstructure.}\label{fig1}
\end{center}
\end{figure}

The OLC nanoparticles were produced by the detonation method of precursor nanodiamonds, 
as described in Ref.~\onlinecite{Kuznetsov1994}. The particles were extracted from the
detonation soot by oxidative removal of non-diamond carbons using a
hot mixture of acids. OLC materials were produced by annealing of nanodiamond powder in vacuum
at $1650$ $^\circ$C for $3$ h. As shown in TEM images of Figs. \ref{fig1}(a) and \ref{fig1}(b), the OLC particles were
arranged in aggregates of nearly toughing particles, and the primary particle
size of the OLCs ranged from $\sim 4$ nm to $\sim 7$ nm,  In the preparation of the OLC-PMMA composites,
OLCs were first heat treated in a furnace at $400$ $^\circ$C for $2$ hours, in order to eliminate
the graphitic layers present between the onions and to functionalize their surface with oxygen-containing groups to improve 
the dispersion in the PMMA matrix. In the next
step, isopropyl alcohol (IPA) was used for the pre-dispersion of the OLCs. A few drops of IPA (enough
to wet the OLC powder) were added to the OLC powder and the suspension was sonicated in a
bath for $1$ hour. The PMMA solution was added immediately afterwards. The polymer-OLC
mixture was stirred for 1 hour followed by probe sonication for $15$ minutes. The obtained ink
was deposited on a glass slide and baked on a hotplate at 100 $^\circ$C for $30$ minutes. The samples
were kept at 50 $^\circ$C overnight to assure the complete polymerization of the composite. 
We have prepared samples with OLC concentrations of $x=0.5$, $1$, $2$, and $4$ wt\% with respect
to the weight of PMMA. The corresponding values of the OLC volume fraction are given by
 $\phi=x\rho_\textrm{PMMA}/(\rho_\textrm{OLC}+x\rho_\textrm{PMMA})$,
where  $\rho_\textrm{PMMA}=1.18$ gm cm$^{-3}$ 
and $\rho_\textrm{OLC}=2.25$ gm cm$^{-3}$ are the mass density values of the PMMA and OLC, respectively.

The conductivity, $\sigma$, of OLC-PMMA composites was measured using the standard $4$ probe method. The
golden wires were attached to the samples by means of a carbon paste. The typical dimensions
of the samples were about $5$ mm $\times$ $2$ mm $\times$ $100$ mm, however for each sample the geometrical
factor was measured and applied for conductivity measurement. The pressure dependence of
$\sigma$ was measured under hydrostatic pressure up to 2 GPa in a piston cylinder cell, using
Daphne oil $7373$ as pressure transmitting medium.

TEM images of microtone slices of OLC-PMMA composites, Figs. \ref{fig1}(c) and \ref{fig1}(d), show 
that the agglomerates of OLC particles have average sizes of the order of $100$ nm. 
The overall microstructure of the composite
is therefore characterized by a strongly non-homogeneous distribution of the conducting OLC particles 
within the polymer, which gives rise to ramified regions that are rich in OLC particles interdispersed 
with regions of almost pure PMMA,  Fig.~\ref{fig1}(d). This kind of microstructure, which has been
frequently observed in other carbon-based composites, favors hopping of electrons between closely separated 
OLC particles, promoting thus relatively high level of conductivity, $\sigma$, even at very small filler 
concentrations.\cite{Palaimiene2018,Adriaanse1997,Flandin1999} 

\begin{figure}[t]
\begin{center}
\includegraphics[scale=1,clip=true]{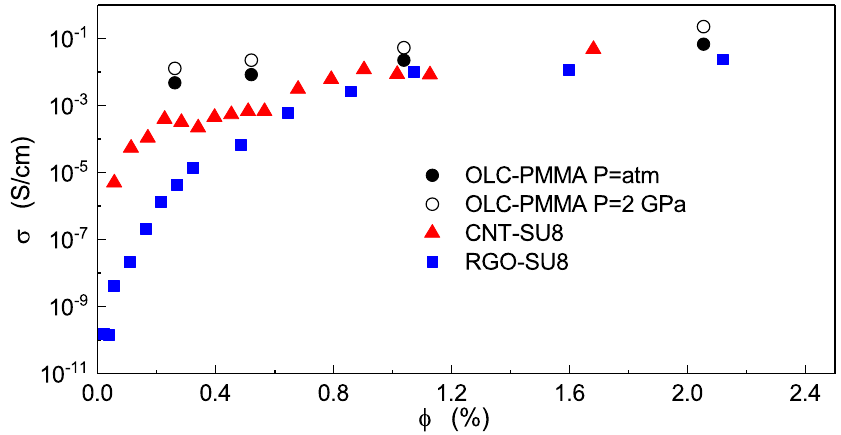}
\caption{Room temperature electrical conductivity of OLC-PMMA composites (filled circles) as a function of the volume fraction $\phi$
of OLC fillers. The conductivity data for the carbon nanotubes (CNT) and reduced graphene oxides (RGO) dispersed in SU8 are taken from 
Refs.~\onlinecite{Grimaldi2013} and \onlinecite{Majidian2014}, respectively.}\label{fig2}
\end{center}
\end{figure}

Enhanced conductivity is indeed measured also in our OLC-PMMA composites, which show values of $\sigma$ that
can be even larger than the conductivity of polymers filled at similar loadings by carbon nanotubes or reduced graphene 
sheets,\cite{Grimaldi2013,Majidian2014} as shown in Fig.~\ref{fig2}. Furthermore, the conductivity of OLC-PMMA decreases by
just one order of magnitude as $\phi$ is reduced from $\approx 2$ \% ($x=4$\%wt) to $\approx 0.25$ \% ($x=0.5$ \%wt)), with
no apparent sign of a percolation transition (see Supplementary Material). Figure~\ref{fig2} shows also that $\sigma$ under an applied
hydrostatic pressure of $P=2$ GPa is almost a factor $3$ larger than the conductivity at atmospheric pressure.
This behavior is consistent with an enhanced probability of tunneling between OLC particles promoted by the reduced 
inter-particle separation under the applied pressure.

\begin{figure*}[t]
\begin{center}
\includegraphics[scale=1,clip=true]{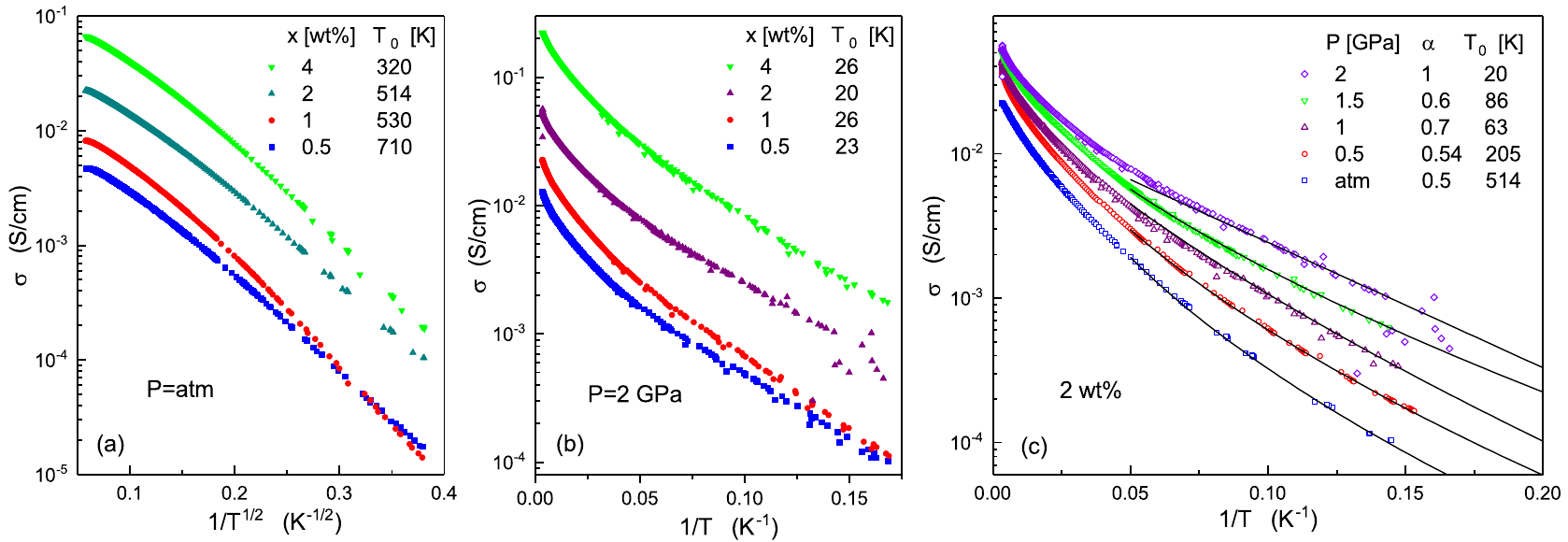}
\caption{Effect of the temperature on the electrical conductivity of OLC-PMMA composites at (a)
ambient pressure and (b) at $P=2$ GPa. Note that in (a) $\sigma$ follows the stretched exponential
behavior of Eq.~\ref{ES} with $\alpha\simeq 1/2$, while the conductivity data at $2$ GPa (b) have activated behavior.
(c): Temperature dependence of the electrical conductivity at different values of the applied
pressure for a sample with $2$ \% wt of OLC particles. The solid lines are fits to Eq.~\eqref{sigmaT} with parameters 
$T_0$ and $\alpha$ given in the legend.}\label{fig3}
\end{center}
\end{figure*}

A more striking effect of the applied pressure is found in the low-temperature behavior of $\sigma$ shown in Fig.~\ref{fig3}. While for all loadings 
and pressure values the conductivity at $T\lesssim 20$ K follows a stretched exponential behavior of the form
\begin{equation}
\label{sigmaT}
\sigma\propto \exp\!\left[-\left(\frac{T_0}{T}\right)^\alpha\right],
\end{equation}
the exponent $\alpha$ changes as a function of $P$. At atmospheric pressure the conductivity is well fitted by $\alpha\simeq 1/2$, Fig.~\ref{fig3}(a), 
indicating a Coulomb gap regime in which hopping of electrons between nearest neighboring OLC particles is hindered by the Coulomb 
interaction.\cite{ES1988,Belo,Pollak} At a pressure of $P=2$ GPa, instead, the low-temperature conductivity follows Eq.~\eqref{sigmaT} with
$\alpha\simeq 1$, Fig.~\ref{fig3}(b), indicating an activated behavior. The change of the exponent from $\alpha\simeq1/2$ to $\alpha\simeq1$ as $P$ 
increases to $2$ GPa is gradual, as shown in Fig.~\ref{fig3}(c) for a sample with $2$ \%wt of OLC particles. Furthermore, the
temperature parameter $T_0$ decreases as $P$ increases. In particular, $T_0$ is of several hundreds Kelvin at ambient pressure
[with some dependence on the OLC concentration, Fig.~\ref{fig3}(a)], while $T_0\approx 20$ K at $P=2$ GPa, with almost no effect of the filler
loading. 

To interpret the experimental results shown in Figs.~\ref{fig2} and \ref{fig3} we express the conductance between a pair $i$ and $j$
of OLC particles as the product of probabilities for tunneling and thermal activation: $G_{ij}=G_0\exp[-2(r_{ij}-D)/\xi-E_{ij}/K_\textrm{B}T]$, 
where $r_{ij}$ is the distance between the centers of two OLCs approximated by identical spherical conductors of diameter $D$, $\xi$ is the electron 
localization length, $K_\textrm{B}$ is the Boltzmann constant, and $E_{ij}$ is the energy difference for hopping between particles $i$ and $j$.
By treating any OLC pair as immersed in an effective dielectric field developed by the PMMA and the other OLC fillers, the energy difference
becomes $E_{ij}\simeq e^2\mathbf{\tilde{C}}^{-1}_{ij}/\tilde{\kappa}$, where $e$ is the electron charge and $\mathbf{\tilde{C}}$ is the capacitance 
matrix of two spherical conductors embedded in the effective dielectric medium with permittivity $\tilde{\kappa}$. 
Following Ref.~\onlinecite{Lekner2011}, we find (see Supplementary Material):
\begin{equation}
\label{C4}
\mathbf{\tilde{C}}^{-1}_{ij}\simeq\left\{
\begin{array}{ll}
1/r_{ij}, & r_{ij}\gg D \\
1/(D\ln 2), & r_{ij}=D
\end{array}\right.,
\end{equation}
which can be more conveniently approximated by $\mathbf{\tilde{C}}^{-1}_{ij}\simeq (\delta_{ij}+D\ln 2)^{-1}$,
where $\delta_{ij}=r_{ij}-D$. In this way, the conductance between a pair of particles reduces to
\begin{equation}
\label{cond1}
G(r_{ij})\simeq G_0\exp\!\left[-\frac{2\delta_{ij}}{\xi}-\frac{e^2}{\tilde{\kappa}(\delta_{ij}+D\ln 2)K_\textrm{B}T}\right].
\end{equation}
Since the composite conductivity $\sigma$ is dominated by the highest pair conductances, and since these are most probably found
within the OLC aggregates, we approximate $\sigma$ by averaging $G(r_{ij})$ over particle pairs within the aggregates: 
\begin{equation}
\label{cond2}
\sigma\propto \sum_{i,j \in\textrm{aggregates}}G(r_{ij})\propto \rho_\textrm{aggr}\int_D^\infty\! dr r^2 P(r)G(r),
\end{equation}
where $\rho_\textrm{aggr}$ is the local number density of OLC particles in the aggregates and $P(r)$ is the corresponding 
pair distribution function of particles.
Similar to what is observed in other ramified aggregated structures like, for example, high-structured carbon-black 
composites\cite{Salome1991,Rieker2000} or some colloidal systems,\cite{Dinsmore2001} we expect $P(r)$ to 
follow for $r>D$ a fractional power-law decay up to a distance, $D_\textrm{aggr}$, of the order of the 
mean size of the aggregates. Here we make the assumption that at low temperatures $P(r)$ has a much weaker $r$-dependence 
than $G(r)$, so as to replace $P(r)$ by a constant, $\chi$, up to distances comparable to $D_\textrm{aggr}\approx 100$ nm:
\begin{equation}
\label{cond3}
\sigma\propto\frac{\phi_\textrm{aggr}\chi}{D}\int_0^{D_\textrm{aggr}}\! d\delta \exp\!\left[-\frac{2\delta}{\xi}-\frac{e^2}{\tilde{\kappa}(\delta+D\ln 2)K_\textrm{B}T}\right],
\end{equation}
where the local volume fraction within the region of aggregates, $\phi_\textrm{aggr}=\pi \rho_\textrm{aggr}D^3/6$,
can be considerably larger than the macroscopic volume fraction $\phi$,\cite{Doyle1990} which explains the high conductivity
values reported in Fig.~\ref{fig2}.

At low temperatures the integral in Eq.~\eqref{cond3} is dominated
by the largest value of the integrand function. We obtain therefore that $\sigma$ follows the Coulomb gap regime
\begin{equation}
\label{ES}
\sigma\propto e^{-\sqrt{T_\textrm{ES}/T}},\,\,\,\, T_\textrm{ES}=\frac{8 e^2}{\xi\tilde{\kappa}K_\textrm{B}},
\end{equation} 
when the optimal hopping distance that maximizes the integrand,
\begin{equation}
\label{deltastar}
\delta^*=\sqrt{\frac{\xi e^2}{2\tilde{\kappa}K_\textrm{B}T}}-D\ln\!2,
\end{equation}
is positive, while for $\delta^*<0$ the conductivity follows an Arrhenius law:
\begin{equation}
\label{TA}
\sigma\propto e^{-T_\textrm{A}/T},\,\,\,\,T_\textrm{A}=\frac{e^2}{\tilde{\kappa}K_\textrm{B}D\ln\!2},
\end{equation}
because in this case the integrand is a monotonous decreasing function of $\delta$ in the range  $(0,D_\textrm{aggr})$, and
so its the largest value is at $\delta=0$. 

Equations \eqref{ES}-\eqref{TA} suggest that in OLC-PMMA composites the cross-over of $\sigma$ from the Coulomb gap regime to the activated behavior
is related to the sign change of $\delta^*$, from positive to negative, induced by the applied pressure. We claim that such sign change is possibly driven 
by the response of the effective permittivity to $P$. Indeed, since the OLC particles are essentially
rigid with respect to the much softer PMMA matrix, at a given temperature the main effect of the pressure is that of enhancing the concentration 
of OLC particles, thus increasing $\tilde{\kappa}$. Larger values of $\tilde{\kappa}$ may in turn make $\delta^*$ smaller than its value at atmospheric pressure,
or even negative.

To see if this mechanism is consistent with our experimental results, we first consider the Coulomb gap regime observed at the atmospheric pressure.
The values of $T_0$ reported in Fig.~\ref{fig3}(a) range from $\sim 300$ K to $\sim 700$ K as the OLC concentration is reduced. We take $T_0\approx 500$ K
as a representative value and we identify it with $T_\textrm{ES}$. From Eq.~\eqref{ES} we obtain $\xi\tilde{\kappa}_0\approx 270$ nm, where 
$\tilde{\kappa}_0$ is the effective permittivity at atmospheric pressure. Since $\delta^*$ has to be positive to give rise to
the Coulomb gap regime, we obtain from Eq.~\eqref{deltastar} $\xi>4D\ln\!2\sqrt{T/T_\textrm{ES}}$. Using $D\approx 5$ nm and $T=10$ K this
condition leads to $\xi>2$ nm and, consequently, $\tilde{\kappa}_0<135$. 

Let us now consider the conductivity at $P=2$ GPa of Fig.~\ref{fig3}(b), which follows 
an activated regime with $T_\textrm{A}\simeq 20$ K. From Eq.~\eqref{TA} we get $\tilde{\kappa}_1=e^2/T_\textrm{A}K_\textrm{B}D\ln\!2\simeq 240$, where now $\tilde{\kappa}_1$
is the value of $\tilde{\kappa}$ at $2$ GPa. Since an activated behavior comes into play only if $\delta^*$
is negative, for $T=10$ K we obtain from Eq.~\eqref{deltastar} $\xi<2(T/T_\textrm{A})D\ln\!2 \approx 3.5$ nm.

From the above analysis, therefore, we conclude that the two different regimes of $\sigma$ reported in Figs.~\ref{fig2}(a) and \ref{fig2}(b) can be
explained by the increase of the effective permittivity upon applied pressure, from less than about $135$ at atmospheric pressure to
about $240$ at $2$ GPa, assuming a localization length comprised between $\sim 2$ nm and $\sim 3.5$ nm. Although this range of $\xi$ values is
quite realistic, an effective permittivity of order $100$ appears to be far larger than that measured in other OLC composites at comparable
filler loadings.\cite{Macutkevic2009,Kranauskaite2018} It should be noted, however, that within the regions of the OLC aggregates, $\tilde{\kappa}$ 
could largely exceed the macroscopic effective permittivity because of the enhanced local volume fraction of the conducting 
fillers.\cite{Doyle1990,Starke2006} 
To see this, we adopt the cluster model of Ref.~\onlinecite{Doyle1990} and express the local permittivity in the aggregates as 
\begin{equation}
\label{kappa}
\tilde{\kappa}\simeq\kappa_0\frac{3\phi_c}{(2-\phi_c)(\phi_c-\phi_\textrm{aggr})},
\end{equation}
where $\kappa_0\simeq 3$ is the permittivity of PMMA and $\phi_c$ is the critical packing fraction at which the effective polarization of the aggregate is one.\cite{Doyle1990}
Since we are considering aggregates of metallic spheres, $\phi_c$ can be identified as the minimal volume fraction such that a cluster of touching spheres spans the
entire aggregate. Quite intuitively, Eq.~\eqref{kappa} shows that large values of $\tilde{\kappa}$ can be attained for $\phi_\textrm{aggr}$ sufficiently close 
to $\phi_c$, and that small variations of $\phi_\textrm{aggr}$ can induce significant changes in the effective permittivity. By identifying $\phi_c$ as the volume fraction
of a random loose packing arrangement of spheres ($\simeq 0.55$), we obtain from Eq.~\eqref{kappa} that our estimate $\tilde{\kappa}<135$ obtained at atmospheric pressure
is reproduced by $\phi_\textrm{aggr}< 0.534$. Furthermore, the relative change of the local permittivity under the application of $2$ GPa, $(\tilde{\kappa}_1-\tilde{\kappa}_0)/\tilde{\kappa}_0>79$ \%, 
can be explained by a relative enhancement of only $\gtrsim 2$ \% in the local volume fraction of the aggregates.

In conclusion, we have reported conductivity measurements of OLC-PMMA composites as a function of volume fraction of OLC fillers, temperature
and applied pressure up to $2$ GPa. Our main finding is that the low-temperature of $\sigma$ can be changed from a Coulomb
gap regime at atmospheric pressure to an activated behavior at $2$ GPa. We interpret this feature by the increase of the effective permittivity upon
the applied pressure, which reduces the optimal hopping distance at low temperatures.

See Supplementary Material for a percolation analysis of the conductivity data and a derivation of the interaction Coulomb term.

We thank Dr. A. Pisoni, Dr. S. I. Moseenkov and Prof. A. I. Romanenko for their technical assistance. The Swiss National Science Foundation supported this work
(Grant Nos. 200020-163441 and 200021-140557).

\renewcommand{\thefigure}{S\arabic{figure}}
\renewcommand{\theequation}{S\arabic{equation}}
\renewcommand{\thetable}{S\arabic{table}}
\setcounter{equation}{0}
\setcounter{figure}{0}
\setcounter{table}{0}

\section{Electrical transport in onion-like carbon-PMMA - nanocomposites: Supplementary Material}

\section{Percolation analysis of the conductivity}
In conductor-insulator composites it is often assumed that the conductivity follows a percolation behavior
as a function of the conducting filler volume fraction $\phi$ characterized by the power-law behavior:
\begin{equation}
\label{power}
\sigma\simeq \sigma_0 (\phi-\phi_c)^t,
\end{equation}  
where $\sigma_0$ is a conductivity prefactor, $\phi_c$ is the critical volume fraction of the conducting phase, and $t$
is the conductivity exponent. Equation \eqref{power} follows from the assumption that electrical connections are established
between conducting fillers (the OLC particles in our case) in such a way that the conductances between any two particles are either 
nonzero, when the relative particle separation is small enough, or zero when the particles are far apart. The bulk conductivity of the composite
exceeds therefore the one of the matrix if there exists a macroscopic cluster of connected particles spanning the entire system,
that is, if $\phi>\phi_c>0$. Although the value of $\phi_c$ depends on the type of distribution of the conducting fillers, the value of 
the exponent is universal and is approximately equal to $2$ in three dimensional systems.

\begin{figure}[t]
	\begin{center}
		\includegraphics[scale=0.35,clip=true]{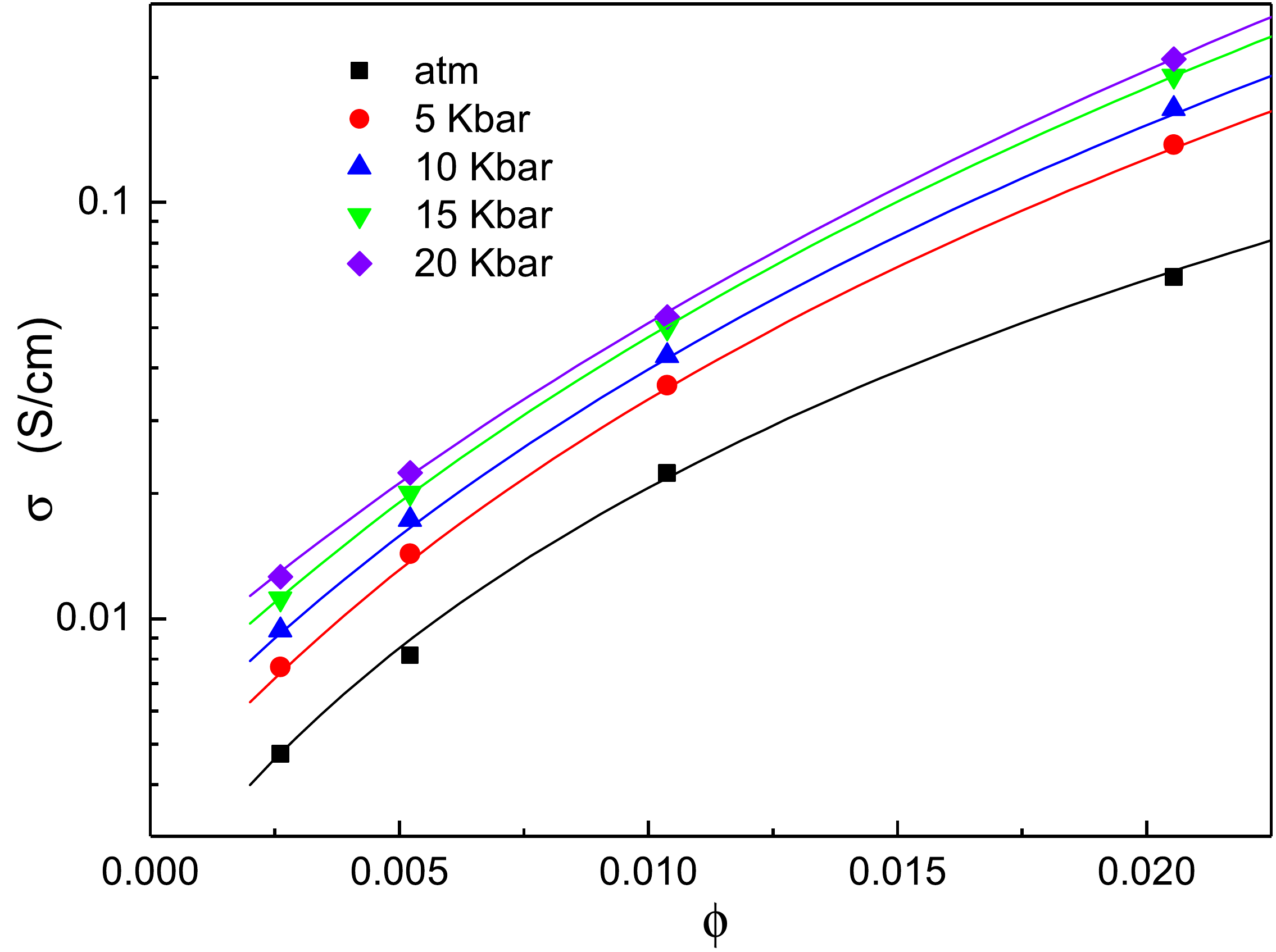}
		\caption{Room temperature electrical conductivity of OLC-PMMA composites (symbols) as a function of OLC volume fraction for 
			different applied pressures. The solid lines are the best fits to Eq.~\eqref{power}. The fitting parameters are reported in
			Table~\ref{table1}.}\label{figS1}
	\end{center}
\end{figure}

Figure~\ref{figS1} shows the conductivity measured in our OLC-PMMA composites for different values of the hydrostatic pressure $P$. 
Although Eq.~\ref{power} formally fits the measured $\sigma$ for all values of $P$, the resulting best fitting values for  $\phi_c$ and $t$ 
reported in Table \ref{table1} are unphysical. We find indeed that not only the transport exponent $t$ is systematically larger or much larger than 
universal value $t\simeq 2$ but, even more strikingly, that the percolation threshold $\phi_c$ is always negative. 
These results clearly indicate that percolation of the OLCs particles is inadequate in describing the conductivity of OLC-PMMA composites.
This is confirmed by the low-temperature conductivity measurements reported in the main text, which show that $\sigma$ is proportional
to $\exp[-(T_0/T)^\alpha]$, a clear indication that $\sigma$ is governed by electron hopping processes between the OLC particles and that
there is not a cut-off in the inter-particle conductances.

\begin{table}[b]
	\caption{Values of $\phi_c$ and $t$ that best fit the measured conductivity of OLC-PMMA composites at room temperature.}
	\label{table1}
	\begin{ruledtabular}
		\begin{tabular}{ccc}
			$P$ (Kbar) & $\phi_c(\%)$ & $t$ \\
			\hline
			$0$ &  $-0.6\pm 0.3$ & $2.4\pm 0.6$ \\
			$5$ &  $-1\pm 0.1$ & $3.3\pm 0.2$ \\
			$10$ &  $-1.3\pm 0.2$ & $3.6\pm 0.4$ \\
			$15$ & $-1.4\pm 0.2$ & $4.0\pm 0.3$ \\
			$20$ &  $-2.0\pm 0.6$ & $5\pm 1$ 
		\end{tabular}
	\end{ruledtabular}
\end{table}

\section{Derivation of the interaction term}

We approximate the OLC particles by spherical conductors of identical diameter $D$. The energy difference for hopping between 
two particles $i$ and $j$ is $E_{ij}=e^2\mathbf{C}^{-1}_{ij}$, where $e$ is the electron charge and $\mathbf{C}$ is the capacitance matrix 
for a system of conducting OLC particles embedded in the polymer.
We assume that $E_{ij}$ can be approximated by the Coulomb interaction between two particles immersed in an effective dielectric field
developed by the PMMA and the other OLC fillers. In this way write $E_{ij}=e^2\mathbf{\tilde{C}}^{-1}_{ij}/\tilde{\kappa}$, where $\tilde{\kappa}$ is 
the effective permittivity and
\begin{equation}
\label{matrix}
\mathbf{\tilde{C}}=\left(
\begin{array}{cc}
C_d & C_o \\
C_o & C_d
\end{array}\right)
\end{equation}
is the capacitance matrix of two isolated conductor embedded in the effective dielectric. For distances $r_{ij}$ between the particle centers much 
larger than $D$  the diagonal and off-diagonal elements
of $\mathbf{\tilde{C}}$ reduce respectively to $C_d=D/2$ and $C_o\simeq -(D/2)^2/r_{ij}$, while in the limit of close approach, $r_{ij}\approx D$, 
they are approximated by:\cite{Lekner2011}
\begin{equation}
\label{C2}
C_d\simeq \frac{D}{4}\left(\ln\sqrt{\frac{8D}{\delta_{ij}}}+\gamma\right),\,\,\,\,
C_o\simeq -\frac{D}{4}\left(\ln\sqrt{\frac{D}{2\delta_{ij}}}+\gamma\right),
\end{equation}
where $\gamma\simeq 0.5772$ is the Euler constant and $\delta_{ij}=r_{ij}-D$ is the distance between the surfaces of the two particles. 
From the inversion of Eq.~\eqref{matrix} and making use of the limiting values of $C_d$ and $C_o$ at $r_{ij}\gg D$
and $r_{ij}=D$ we obtain
\begin{equation}
\label{C4}
\mathbf{\tilde{C}}^{-1}_{ij}=-\frac{C_o}{C_d^2-C_o^2}\simeq\left\{
\begin{array}{ll}
1/r_{ij}, & r_{ij}\gg D \\
1/(D\ln 2), & r_{ij}=D
\end{array}\right.
\end{equation}

\end{document}